\documentclass[preprint,12pt,longbibliography,eqsecnum,preprintnumbers,nofootinbib]{revtex4}
\renewcommand{\baselinestretch}{1.4}
\usepackage[bookmarks=true,colorlinks,linkcolor=red,urlcolor=blue,citecolor=blue]{hyperref}
\usepackage[usenames]{color}
\usepackage{dcolumn}
\usepackage{feynmf}

\usepackage[normalem]{ulem}
\usepackage{amsmath}
\usepackage{enumerate}
\usepackage{epsfig}
\usepackage{yfonts}

\usepackage{bm}

\usepackage{amsmath,bm}
\usepackage{amssymb}
\usepackage{graphicx}
\usepackage{amsfonts}         
\usepackage{fancybox}   
\usepackage{yfonts}

\def\comments#1{}

\def\bra#1{{\langle}#1|}
\def\ket#1{|#1\rangle}

\def\vev#1{\langle{#1}\rangle}

\def\CN{{\cal N}}
\def\CO{{\cal O}}

\def\II{\relax{I\kern-.10em I}}

\def\IB{\relax{\rm I\kern-.18em B}}
\def\ID{\relax{\rm I\kern-.18em D}}
\def\IE{\relax{\rm I\kern-.18em E}}
\def\IF{\relax{\rm I\kern-.18em F}}
\def\IG{\relax\hbox{$\inbar\kern-.3em{\rm G}$}}
\def\IGa{\relax\hbox{${\rm I}\kern-.18em\Gamma$}}
\def\II{\relax{\rm I\kern-.18em I}}
\def\IK{\relax{\rm I\kern-.18em K}}

\def\inbar{\,\vrule height1.5ex width.4pt depth0pt}

\def\frac#1#2{{#1 \over #2}}

\newdimen\tableauside\tableauside=1.0ex
\newdimen\tableaurule\tableaurule=0.4pt
\newdimen\tableaustep
\def\phantomhrule#1{\hbox{\vbox to0pt{\hrule height\tableaurule width#1\vss}}}
\def\phantomvrule#1{\vbox{\hbox to0pt{\vrule width\tableaurule height#1\hss}}}
\def\sqr{\vbox{%
  \phantomhrule\tableaustep
  \hbox{\phantomvrule\tableaustep\kern\tableaustep\phantomvrule\tableaustep}%
  \hbox{\vbox{\phantomhrule\tableauside}\kern-\tableaurule}}}
\def\squares#1{\hbox{\count0=#1\noindent\loop\sqr
  \advance\count0 by-1 \ifnum\count0>0\repeat}}
\def\tableau#1{\vcenter{\offinterlineskip
  \tableaustep=\tableauside\advance\tableaustep by-\tableaurule
  \kern\normallineskip\hbox
    {\kern\normallineskip\vbox
      {\gettableau#1 0 }%
     \kern\normallineskip\kern\tableaurule}%
  \kern\normallineskip\kern\tableaurule}}
\def\gettableau#1 {\ifnum#1=0\let\next=\null\else
  \squares{#1}\let\next=\gettableau\fi\next}

 \def\eqnn#1{\xdef #1{(\secsym\the\meqno)}\writedef{#1\leftbracket#1}%
 \global\advance\meqno by1\wrlabeL#1}
 \def\eqna#1{\xdef #1##1{\hbox{$(\secsym\the\meqno##1)$}}
 \writedef{#1\numbersign1\leftbracket#1{\numbersign1}}%
 \global\advance\meqno by1\wrlabeL{#1$\{\}$}}
 \def\eqn#1#2{\xdef #1{(\secsym\the\meqno)}\writedef{#1\leftbracket#1}%
 \global\advance\meqno by1$$#2\eqno#1\eqlabeL#1$$}

\global\newcount\itemno \global\itemno=0

\def\itemaut#1{\global\advance\itemno by1\noindent\item{\the\itemno.}#1}

\def\({\left(}
\def\){\right)}

\def\eg{{\it e.g.}}
\def\ie{{\it i.e.}}

\newif{\ifeq}           
\eqtrue                 
\newcommand{\be}{\begin{equation}}
\newcommand{\ee}{\end{equation}}
\newcommand{\bea}{\begin{eqnarray}}
\newcommand{\eea}{\end{eqnarray}}
\newcommand{\bean}{\begin{eqnarray*}}
\newcommand{\eean}{\end{eqnarray*}}

\def\({\left(}
\def\){\right)}
\def\[{\left[}
\def\]{\right]}

\renewcommand{\O}{{\cal O}}

\def\CO{\O}

\newcommand{\IR}{{\mathbb R}}

\def\ie{{\it i.e.}}

\newcommand{\lsim}{\,\raise.3ex\hbox{$<$\kern-.75em\lower1ex\hbox{$\sim$}}\,}
\newcommand{\gsim}{\,\raise.3ex\hbox{$>$\kern-.75em\lower1ex\hbox{$\sim$}}\,}

\newif{\ifeq}
\eqtrue

\numberwithin{equation}{section}

\def\XX{{\cal X}}
\def\PP{{\cal P}}

\def\aaa{{\mathfrak a}}

\begin{document}

\begin{titlepage}

\begin{flushright}
MIT-CTP/4298
\end{flushright}
\vfil

\begin{center}
{\huge Exactly Stable Collective Oscillations in Conformal Field Theory}\\
\end{center}
\vfil
\begin{center}
{\large Ben Freivogel, John McGreevy and S. Josephine Suh}\\
\vspace{1mm}
Center for Theoretical Physics, MIT,
Cambridge, Massachusetts 02139, USA\\
\vspace{3mm}
\end{center}

\vfil

\begin{center}

{\large Abstract}
\end{center}

\noindent
Any conformal field theory (CFT) on 
a sphere supports completely undamped collective oscillations.
We discuss the implications of this fact for studies of thermalization using AdS/CFT.
Analogous oscillations occur in Galilean CFT,
and they could be observed in experiments on ultracold fermions.

\vfill
\begin{flushleft}
September 2011; revised December 2011
\end{flushleft}
\vfil
\end{titlepage}

\renewcommand{\baselinestretch}{1.1}  

\renewcommand{\arraystretch}{1.5}



\section{Introduction}

Conformal field theories (CFTs) are interesting for many reasons:
they arise in the study of critical phenomena, on
 the worldsheet of fundamental strings, and in the holographic dual of anti-de Sitter spacetimes
\cite{Maldacena:1997re, Gubser:1998bc,  Witten:1998qj}.  It is plausible that all quantum field theories are relevant deformations of  conformally invariant ultraviolet fixed points.


Here we describe an exotic property of any CFT in any number of dimensions.
Any relativistic CFT whose spatial domain is a sphere contains a large class of non-stationary states 
whose time evolution is periodic, with frequencies that are integer multiples of the inverse radius of the sphere.  

Nonrelativistic CFTs that realize the Schr\"odinger algebra also support undamped oscillations in the presence of a spherically symmetric harmonic potential.
Cold fermionic atoms with tuned two-body interactions can provide
an experimental realization of such a system, 
and the modes we discuss (which, in this context, 
have been discussed previously in \cite{Castin2004})
could be (but have not yet been) observed.

The existence of these permanently oscillating many-body states is guaranteed by $sl(2,\IR)$ subalgebras of the conformal algebra, formed from the Hamiltonian and combinations of operators which act as ladder operators for energy eigenstates.  
It is striking that any CFT in any dimension, regardless of the strength or complication of its interactions,
has states that undergo {\it undamped} oscillation.

We emphasize the distinction between these oscillating states and an energy eigenstate of an arbitrary Hamiltonian.
Time evolution of an energy eigenstate is just multiplication of the wavefunction by a phase -- nothing happens.
Given knowledge of the exact energy eigenstates of any system, one 
may construct special operators whose correlations oscillate in time in certain states.
In contrast, we show below that in the states described here, accessible physical quantities such as the energy density
vary in time (at leading order in $N^2$ in holographic examples).
Further, the oscillations arise and survive at late times starting from generic initial conditions\footnote
{We explain the precise sense in which they are generic in~\ref{sec:charges}.}.

The persistence of these oscillations conflicts with conventional expectations for thermalization of an 
interacting theory. The conventional wisdom is that an arbitrary initial state will settle down to an equilibrium stationary configuration characterized by its energy and angular momenta (for a recent
discussion of this expectation, see \cite{Hubeny:2011hd}). This is not the case for CFT's due to the presence of these undamped oscillations.

The existence of these modes is due to the existence of extra conserved charges 
in CFT generated by conformal Killing vectors.
Any conserved charge will partition
the Hilbert space of a system into sectors
which do not mix under time evolution.
However, in the conventional situation there will be be a stationary state for each value of the charges which represents equilibrium in that sector.
The states we describe do not approach a stationary state; rather, the {\it amplitude} of the oscillations
is a conserved quantity, as we explain in Section \ref{sec:charges}.

Any system with a finite number of degrees of freedom will exhibit {\it quasiperiodic} evolution: 
the time dependence of \eg~a correlation function in such a system is inevitably a sum of a finite number of Fourier modes.
The class of oscillations we describe can be distinguished from such generic behavior in two respects: 
First, for each oscillating state the period is fixed to be an integer multiple of the circumference of the sphere.
Second, and more importantly, 
the oscillations persist and remain undamped in a particular thermodynamic limit
-- namely, a large-$N$ limit where the number of degrees of freedom per site
diverges.
Many such CFTs are described by classical gravity theories in
asymptotically anti-de Sitter spacetime (AdS).

The latter fact points to a possible obstacle in studying thermalization of CFTs on a sphere using holography: 
when subjecting the CFT to a far-from-equilibrium process, if one excites such a mode of oscillation,
this excitation will not go away, even at infinite $N$.
(The simplest way to circumvent this obstable is to study CFT on the plane.  Then this issue does not arise,
as the frequency of the modes in question goes to zero.)

In holographic theories, certain oscillating states with period equal to the circumference of the sphere have an especially simple bulk description. Begin with a Schwarzschild-AdS black hole in the bulk. This is dual to a CFT at finite temperature on the boundary. Now boost the black hole. This boost symmetry is an exact symmetry of AdS, and it creates a black hole that ``sloshes" back and forth forever\footnote
{Such oscillations are known to S.~Shenker and L.~Susskind as ``sloshers".
The action of conformal boosts on AdS black holes has been studied previously 
in \cite{Horowitz:1999gf}.
}. The dual CFT therefore has periodic correlators.
 The boosted black hole is related to the original black hole by a large diffeomorphism that acts nontrivially on the boundary. The boundary description of the boosting procedure is to act with a conformal Killing vector on the thermal state. The relevant CKV's are periodic in time, and they act nontrivially on the thermal state, so they produce oscillations.

This document is organized as follows:
In section \ref{sec:osc} we construct the oscillating states explicitly.
In section \ref{sec:charges} we describe conserved charges associated with conformal symmetry whose nonzero expectation value diagnoses oscillations in a given state.
We also derive operator equations 
which demonstrate that 
certain $\ell=1$ moments of the stress-energy tensor
in CFT on a sphere
behave like harmonic oscillators.
In section \ref{sec:holography} we 
specialize to holographic CFTs and 
discuss the gravitational description of a specific class of oscillations around thermal equilibrium.
In section \ref{sec:correlators}, we discuss the effect of the existence of oscillating states on correlation functions of local operators.
We end with 
some explanation of our initial motivation for this work
and some comments and open questions.
Appendix \ref{app:algebra} summarizes the conformal algebra.
Appendix \ref{app:norms} gives some details on the normalization
of the oscillating states.
Appendix \ref{app:goldstone} extends Goldstone's theorem to
explain the linearized oscillations of smallest frequency.
Appendix \ref{app:qnm} constructs one of the consequent linearized modes of
the large AdS black hole, following a useful
analogy with the translation mode of the Schwarzchild black hole in flat space.
Appendix \ref{app:bbh} constructs finite oscillations of an AdS black hole
and calculates observables in the dual CFT.
In the final Appendix \ref{app:galilean}, we explain that 
an avatar of these modes has already been studied in experiments 
on cold atoms, and that it is in principle possible 
to demonstrate experimentally the precise analog of these oscillating states.

\section{Stable oscillations in relativistic CFT on a sphere}
\label{sec:osc}

Consider a relativistic CFT on a $d$-sphere cross time. 
We set $c=1$ and measure energies in units of the inverse radius of the sphere, $R^{-1}$.
The conformal algebra acting on the Hilbert space of the CFT contains
$d$ (non-independent) copies of the $sl(2, \mathbb{R})$ algebra,
\be [H,L_{+}^{i}]=L_{+}^{i}, ~~[H,L_{-}^{i}]=-L_{-}^{i}, ~~[L_{+}^{i}, L_{-}^{i}]=2H \ee
(with no sum on $i$ in the last equation)
as reviewed in Appendix~\ref{app:algebra}.
Here $i = 1...d$ is an index labeling directions in the $\mathbb{R}^d$ in which
$S^{d-1}$ is embedded as the unit sphere.

We construct the oscillating states in question using the $sl(2,\mathbb{R})$ algebras as follows. Let $\ket{\varepsilon}$ be an eigenstate of $H$, $H\ket{\varepsilon}=\varepsilon\ket{\varepsilon}$.
Consider a state of the form 
\be
\label{eq:oscstate}
\ket{\Psi (t=0)}
= \CN e^{\alpha L_{+}+\beta L_{-}}\ket{\varepsilon},
\ee
where $L_{+}= L_{+}^i$ and $L_{-}=L_{-}^j$ for some $i,j \in \{1,...,d\}$, $\CN$ is a normalization constant
given in Appendix \ref{app:norms}, and $\alpha$ 
and $\beta$ are complex numbers.
If $\alpha, \beta$ are chosen so that the operator in \eqref{eq:oscstate} is unitary,
there is no constraint on $\alpha, \beta$ from normalizability;
more generally, finite $\CN$ constrains $\alpha, \beta$ as described in Appendix \ref{app:norms}.
It evolves in time as\footnote{This can be checked using the Baker-Campbell-Hausdorff formula $e^{B}Ae^{-B}=\sum\limits_{n=0}^{\infty}\frac{1}{n!}(ad_{B})^{n}A$.}
\be
\label{eq:evolution}
\ket{\Psi(t)}=\CN e^{-iHt}e^{\alpha L_{+}+\beta L_-}\ket{\varepsilon}= \CN e^{\alpha e^{-it}L_{+}+\beta e^{it}L_{-}}e^{-iHt}\ket{\varepsilon}= \CN e^{-i\varepsilon t}e^{\alpha(t)L_{+}+\beta(t)L_{-}}\ket{\varepsilon},
\ee
where $\alpha(t)=\alpha e^{-it}$ and $\beta(t)=\beta e^{it}$. 
$\ket{\Psi}$ has been constructed in analogy with coherent or squeezed states in a harmonic oscillator, where one also has ladder operators for the Hamiltonian. Restoring units, $\ket{\Psi}$ is seen to oscillate with frequency $1/R$.

More generally, the time evolution of a state 
\be
\label{eq:general}
\ket{\Psi(t=0)}=g(L_+^1,...,L_+^d,L_-^1,...,L_-^d)\ket{\varepsilon}
\ee
where $g$ is any regular function of the $2d$ variables $L_+^i, L_{-}^j$, $1\leq i,j\leq d$, is given by replacing $L_\pm^i$ with $L_\pm^i e^{\pm i t}$ $\forall$ $i$. 
For example, 
oscillating state with frequency $2/R$
is given by \eqref{eq:general} with 
$g=\sum_{m=0}^\infty {1\over \(m!\)^2} \(\alpha L_{+}^{2}+\beta L_{-}^{2} \)^m $, 
where $L_{+}$ can be any of $L_{+}^{i}$, $1\leq i \leq d$, and similarly for $L_{-}$.\footnote{
The coefficient $\(m!\)^2$ is designed to give the state a finite norm.
We note that a state of the form \eqref{eq:general} 
with $g=e^{\alpha L_{+}^{n}+\beta L_{-}^{n}}$ is only normalizable for $n = 0, 1$;
in particular, the direct analog of a squeezed state does not seem to exist.}
Note, however, that not every such function $g$ produces an oscillating state - a state in which physical quantities such as energy density display oscillation - as opposed to a state whose time evolution is given by a trivial but nonetheless periodic factor, as with an energy eigenstate. Take for example $g=L_{+}$, which merely produces another eigenstate. The criterion for oscillation which we establish in section~\ref{sec:charges} below ($\langle Q^{\pm,i}\rangle \neq 0$ for some $i$) can be applied to an initial state $\ket{\Psi(t=0)}=g\ket{\varepsilon}$ to determine whether $\ket{\Psi}$ is indeed an oscillating state. In addition, normalizability of $\ket{\Psi}$ will constrain the function $g$ in a manner which we have not determined.

To retain the simple time evolution in the above construction, the stationary state $\ket{\varepsilon}$ cannot be replaced by a nonstationary state
$\sum\limits_{i}c_i \ket{\varepsilon_{i}}$, with terms that evolve in time with distinct phases. However, it can be replaced by a stationary density matrix $\rho$, which evolves in time with a phase $A$ (possibly zero), 
\be [H, \rho] = A \rho. \ee
Examples include an energy eigenstate ${\rho} = \ket{\varepsilon}\bra{\varepsilon}$,  and thermal density matrices $\rho = e^{ - \beta H }Z_{\beta}^{-1}$.
Given such a $\rho$, an initial ensemble
\begin{align}
\tilde{\rho}(t=0)=\CN e^{\alpha L_{+}+\beta L_{-}}\rho e^{\beta^* L_{+}+\alpha^* L_{-}}
\end{align}
evolves in time as 
\be \label{eq:collective} \tilde{\rho}(t)=\CN e^{-i A t}e^{\alpha(t) L_{+}+\beta(t) L_{-}}\rho e^{\beta(t)^* L_{+}+\alpha(t)^* L_{-}}.\ee

We refer to the collective oscillation $\tilde{\rho}$ as having been ``built on" $\rho$, in the same way  $\ket{\Psi}$ was built on $\ket{\varepsilon}$, by acting with a sufficiently constrained function of ladder operators $g$ at $t=0$. In section \ref{sec:holography}, we give a holographic description of the subset of oscillations built on a thermal density matrix for which $g=e^{\alpha L_{+}+ \beta L_{-}}$ and $\alpha L_{+}+ \beta L_{-}$ is anti-Hermitian.

\section{Diagnosing the Amplitude of Oscillations}
\label{sec:charges}

It may be useful to be able to diagnose the presence of the above oscillations in a generic state of a CFT. Here we identify conserved charges associated with conformal Killing vectors, which correspond to the conserved amplitudes of possible oscillations.  

Consider CFT on a spacetime $M$. 
Given a conformal Killing vector field (CKV) $\xi^{\mu}$ on $M$, there is an associated current and charge
acting in the Hilbert space of the CFT
\be
j_{\xi}^{\mu}=T^{\mu\nu}\xi_{\nu}, ~~~~Q_{\xi}=\int_{\Sigma}d^{d-1}x\sqrt{g}~j^{\mu}_{\xi}n_{\mu},
\ee
where $\Sigma$ is a spatial hypersurface and $n^{\mu}$ is a normal vector. 
$\nabla_{\mu}j_{\xi}^{\mu}$ is given by a state-independent but $\xi$-dependent constant
\be
\label{eq:conserved}
\nabla_{\mu}j_{\xi}^{\mu}=\nabla_{\mu}T^{\mu\nu}+T^{\mu\nu}(\nabla_{\mu}\xi_{\nu}+\nabla_{\nu}\xi_{\mu})=\alpha_{\xi}T^{\mu}_{\mu}, 
\ee
where $ \alpha_{\xi} = {2\over d} \nabla_\mu \xi^\mu$ is a c-number which vanishes when $\xi^{\mu}$ is an exact Killing vector field (KV)\footnote
{Note that in \eqref{eq:conserved} we have assumed $T^{\mu\nu}$ 
is both symmetric and traceless, up to a trace anomaly.}.
Then ${d \over dt} Q_{\xi} = \int_{\Sigma} d^{d-1}x \sqrt{g}~ \alpha_\xi T^\mu_\mu ~ n_t$,
which vanishes for CFT on $\mathbb{R}\times S^{d-1}$
as follows.
The trace anomaly has angular momentum $\ell = 0$.
The proper CKVs have $\ell=1$ and hence the associated $\alpha_\xi$ have $\ell = 1$.  The integral over the sphere vanishes and $Q_{\xi}$ is exactly conserved.

$\xi^{\mu}$'s can be obtained explicitly by projecting $J^{ab}=i(X^{a}\partial^{b}-X^{b}\partial^{a})$, $a,b=-1,0,1,...,d$ in $\mathbb{R}^{(2,d)}$ to the $r\rightarrow \infty$ boundary of $AdS_{d+1}$, the hypersurface $\sum\limits_{i=1}^{d}\Omega^{i\,2}=1$ in coordinates $(r, t, \Omega^{1},...,\Omega^{d})$ with $r\geq0$, $-\infty<t<\infty$, defined by
\begin{align}
X_{-1}=R\sqrt{1+r^2}\cos{t},\\
X_{0}=R\sqrt{1+r^2}\sin{t},\\
X_{i}=R r \Omega_{i}.
\end{align}

In particular, CKVs which are not KVs, corresponding to boosts $J^{\pm,i}\equiv J^{-1,i}\pm i J^{0,i}$ in $\mathbb{R}^{(2,d)}$, are
\begin{align}
\xi^{\pm,i}=\mp i e^{\pm it}~\Omega^{i}\partial_{t}-e^{\pm it}\bigg[(1-\Omega^{i2})\partial_{\Omega^{i}}-\Omega^{i}\sum\limits_{j\neq i}\Omega^{j}\partial_{\Omega^{j}}\bigg]\bigg|_{\sum\limits_{k}\Omega^{k\,2}=1}.
\end{align}
Letting $\bigg[(1-\Omega^{i2})\partial_{\Omega^{i}}-\Omega^{i}\sum\limits_{j\neq i}\Omega^{j}\partial_{\Omega^{j}}\bigg]\bigg|_{\sum\limits_{k}\Omega^{k\,2}=1}\equiv f^{i\alpha}(\theta)\partial_{\theta^{\alpha}}$ where $\theta^{\alpha}$, $\alpha=1,...,d-1$ are coordinates on $S^{d-1}$, the associated charges are
\begin{align}
\label{eq:initialconditions}
Q^{\pm,i}=\mp i e^{\pm it}\int_{S^{d-1}}d^{d-1}\theta\sqrt{g}~T_{tt}\Omega^{i}-e^{\pm it}\int\limits_{S^{d-1}}{d^{d-1}\theta\sqrt{g}~T_{t\alpha}f^{i\alpha}(\theta)}.
\end{align}

Now consider the quantities 
\be \XX^{i}\equiv\int_{S^{d-1}}d^{d-1}\theta\sqrt{g}~T_{tt}\Omega^{i}, 
~~~~~~
\PP^{i}\equiv-\int_{S^{d-1}}{d^{d-1}\theta\sqrt{g}~T_{t\alpha}f^{i\alpha}(\Omega)}~.  \ee
These are the $i$th coordinate of the center of mass of the CFT state in the embedding space of $S^{d-1}$,
and its momentum, respectively.
From conservation of the charges $Q^{\pm,i}$,
they satisfy the operator equations
\begin{align}
\dot{\XX}^{i}-\PP^{i}=0 ~~~~\dot{\PP^{i}}+\XX^{i}=0
\end{align}
-- they undergo simple harmonic oscillation.
From \eqref{eq:initialconditions}, the initial conditions for these oscillations are determined by the conformal charges.

It follows that given an arbitrary state, if any of the $2d$ expectation values $\langle Q^{\pm,i} \rangle$ are non-zero at $t=0$, some of the non-complex, physical quantities  $\langle\XX^{i}\rangle$ and $\langle\PP^{i}\rangle$ will oscillate with undying amplitude.
The fact that $\langle Q^{\pm,i} \rangle \neq 0$ for some $i$ is an open condition
justifies our use of the word `generic' in the Introduction. As a check that the condition is a good criterion for physical oscillation, note that if $\ket{\Psi}$ is an energy eigenstate, $\bra{\Psi} Q^{\pm,i} \ket{\Psi} =0$. This follows from $\bra{\Psi} [H,Q^{\pm,i}]\ket{\Psi} = \bra{\Psi} \pm Q^{\pm,i} \ket{\Psi} = 0$
\footnote{$Q^{\pm,i}=e^{\pm it} L_{\pm}^i$, as explained in following paragraph.} 
using $H\ket{\Psi} = E\ket{\Psi}$. On the other hand, $\bra{\Psi} Q^{\pm,i} \ket{\Psi} =0$ $\forall$ $i$ does not imply that $\ket{\Psi}$ is an energy eigenstate, indicating that there are non-oscillating states which are not  energy eigenstates. Take for example a superposition of energy eigenstates $\ket{\Psi} = \ket{\Psi_1}+\ket{\Psi_2}$, where  $\ket{\Psi_1}$ and $\ket{\Psi_2}$ belong to different towers in the CFT spectrum, where each tower is built on ladder operators acting on a primary energy eigenstate. Then clearly $\bra{\Psi} Q^{\pm,i} \ket{\Psi} =0$ $\forall$ $i$, although $\ket{\Psi}$ is not an eigenstate. 

Finally, we clarify a potentially confusing point.
By construction, all of the ${(d+2)(d+1) \over 2}$ charges $Q^{AB}$ are time-independent.
There is one associated with each generator of the conformal group in $d$ dimensions;
on general grounds of Noether's theorem, they satisfy the commutation relations of so$(2,d)$.
Since the Hamiltonian for the CFT on the sphere $H$ is one of these generators,
and $H$ is not central (\eg\ $[H, L^i_{\pm}] = \pm L^i_{\pm}$),
there may appear to be a tension between the two preceding sentences.
Happily, there is no contradiction: the
time evolution of the conformal charges $Q^{\pm,i}$ arising from their failure to commute with the time-evolution operator
is precisely cancelled by the explicit time dependence of the 
CKVs $\xi^{\pm,i}$:
\be 
\label{eq:charges}
{d \over dt} Q^{\pm,i} = \partial_t Q^{\pm,i} - i [H, Q^{\pm,i}]=0. 
\ee
Thus $Q^{\pm,i} = e^{\pm i t } L_{\pm}^i$.

\section{Holographic Realization of Oscillations}
\label{sec:holography}

The discussion in the previous sections applies to any relativistic CFT on the sphere.  
We now turn to the case of holographic CFTs and the dual gravitational description of a special class of collective oscillations built on thermal equilibrium, for which $g=e^{\alpha L_{+}+ \beta L_{-}}$ and is anti-Hermitian.

\subsection{Bouncing Black Hole}

Consider a relativistic CFT${}_d$ with a gravity dual, on $S^{d-1}$.
We focus on collective oscillations of the form in \eqref{eq:collective},
built on 
the thermal density matrix $\rho=Z_{\beta}^{-1}\sum_{\varepsilon}e^{-\beta \epsilon}\ket{\varepsilon}\bra{\varepsilon}$, with the parameter $\alpha L_{+}+\beta L_{-}$ restricted to be anti-Hermitian.
With this restriction, $e^{\alpha L_{+}+\beta L_{-}} $ is a finite transformation in $SO(2,d)$.

The gravity dual of such a state can be constructed by a `conformal boost' of a black hole, as follows.
Begin with the static global AdS black hole dual to $\rho$ \cite{Witten:1998qj, Witten:1998zw}.
Now consider a non-normalizable bulk coordinate transformation,
which reduces to the finite conformal transformation $e^{\alpha L_{+}+\beta L_{-}}$ at
the UV boundary of AdS.  Such a transformation falls off 
too slowly to be gauge identification, but too fast to change the couplings of the dual CFT;
it changes the state of the CFT.

For example, a coordinate transformation that corresponds to the boost $J_{01}$ in the embedding coordinates $(X_{-1},X_{0},X_{1},...,X_{d})$ of $AdS_{d+1}$, 
maps empty $AdS_{d+1}$ to itself, 
but
will produce a collective oscillation 
when acting on a global AdS black hole\footnote
{The reader may be worried about ambiguities in the procedure of translating a KV on $AdS_{d+1}$ 
into a vector field on the $AdS$ black hole.  
The UV boundary condition and choice of gauge $g_{r\mu}=0$ appears to make this procedure unique;
this is demonstrated explicitly for the linearized modes in appendix \ref{app:qnm}.
}.
The CFT state at some fixed time is of the form 
\be\label{eq:bb} \CN e^{\alpha(L_+^{1}-L_-^{1})}\(\sum_{\varepsilon}e^{-\beta \varepsilon}\ket{\varepsilon}\bra{\varepsilon}\)e^{\alpha^*(L^{1}_{-}-L^{1}_{+})}\ee 
where $\alpha$ is the boost parameter and real, and evolves as in \eqref{eq:collective} with phase $A=0$. In Appendix \ref{app:bbh}, we discuss the bouncing black hole obtained from such a boost of the BTZ black hole.
We 
exhibit oscillations of the stress-energy tensor
expectation value and of entanglement entropy of a subregion
in the corresponding mixed state.

\subsection{Bulk Modes}

In the CFT, excitations corresponding to oscillations of the form in \eqref{eq:bb} are created by modes of the stress-energy tensor -- linear combinations of $L_{+}$ and $L_{-}$ -- 
acting on the thermal ensemble. They can be interpreted as Goldstone excitations resulting from the breaking of conformal symmetry by the thermal state of the CFT. 
We elaborate on this point in Appendix \ref{app:goldstone}. 

For simple large-$N$ gauge theories, these $L_\pm$ are single-trace operators.
The above excitations then translate holographically 
to single-particle modes in the form of solutions to the linearized equation of motion for the metric in the bulk. They have $\ell=1$ on $S^{d-1}$, as all conformal generators including $L_{\pm}$ have $\ell=1$. Their frequency is $\omega=\pm 1/R$ -- oscillations of higher frequency are built with exponentials of higher powers of the stress-energy tensor, and correspond to multi-particle states.

In Appendix \ref{app:qnm}, we construct one such $\ell=1$ mode in the global $AdS_{4}$-Schwarzschild background, from which the others can be obtained by transformations in $SO(2)\times SO(3)$. In general $d$, the $d\cdot 2$ bulk modes in question will be related by symmetries in $SO(2)\times SO(d)$.\footnote{These modes are not compatible with the Regge-Wheeler gauge for the metric \cite{Regge:1957td} common in the literature on quasinormal modes.
They are perturbations of the metric due to coordinate transformations that asymptote to conformal transformations on the boundary; these coordinate transformations do not preserve the Regge-Wheeler gauge. As such, they were not identified in the spectrum of quasinormal modes of 
the metric in $AdS$ black holes studied in the Regge-Wheeler gauge
\cite{Birmingham:2001pj, Cardoso:2001bb, Miranda:2005qx, Starinets:2008fb, Berti:2009kk}.}

Note that in empty global AdS, there are no analogous $\ell=1$ modes, as all global conformal generators annihilate the vacuum in CFT. However, there are single-particle modes which are dual in the CFT to states created by single-trace primary operators acting on the vacuum, and descendant states obtained by 
acting with $L_{+}^{n}$ for some integer $n\geq1$ on such primary states. The latter correspond to linearized oscillations built on energy eigenstates. Together these single-particle modes are a subset of normal modes found in AdS (see \eg~\cite{Berti:2009kk}, \cite{Natario:2004jd}).

\section{Consequence for correlation functions}
\label{sec:correlators}

What does the existence of these modes imply for thermalization 
as measured by correlation functions of local operators?

A generic correlation function $G(t)$ in a thermal state 
will decay exponentially in time
to a mean value associated with Poincar\'e recurrences which is of order $e^{- S}$.
In a large-$N$ CFT, $S \sim T^{d-1} V N^2 $.
This Poincar\'e-recurrence behavior of $G$
is therefore of order $e^{-N^2}$, and does not arise at any order in perturbation theory.
We expect the contributions to generic correlators at any finite order in the $1/N$ expansion
to decay exponentially in time like $e^{- a T t}$ where $a$ is some order-one numerical number.
From the point of view of a bulk holographic description, this is because
waves propagating in a black hole background fall into the black hole;
the amplitude for a particle {\it not} to have fallen into the black hole
after time $t$ should decay exponentially, like $e^{- a T t}$.
At leading order in $N^2$, 
\ie\ in the classical limit in the bulk,
$G(t)$ decays exponentially in time
at a rate determined by the least-imaginary
quasinormal mode of the associated field.
A process whereby the final state at a late time $t \gg 1/T$
is correlated with the initial state is
one by which the black hole retains information about its early-time state,
and hence one which resolves the black hole information problem;
this happens via contributions of order $e^{-N^2}$ \cite{Maldacena:2001kr}.

The preceding discussion of Poincar\'e recurrences is not special to CFT.
In CFT, there are special correlators which are exceptions to these expectations.
A strong precedent for this arises in hydrodynamics, where
correlators of operators which excite hydrodynamic modes
enjoy power-law tails $G_{\text special}(t)\sim {1\over s} t^{-b}$,
where $s$ is the entropy density and $b$ is another number.
In deconfined phases of large-$N$ theories, $s \propto N^2$
and the long-time tails arise as loop effects in the bulk \cite{Kovtun:2003vj}.
For the high-temperature phase of
of large-$N$ CFTs with gravity duals,
the necessary one-loop computation was performed by \cite{CaronHuot:2009iq}.

A similar situation obtains for large-$N$ CFTs on the sphere.
While correlation functions for generic operators have
perturbative expansions in $N$ which decay exponentially in time
and (non-perturbatively) reach the Poincar\'e limit $e^{-N^2}$ at late times,
\be
G_{\text{generic}}(t) \sim \sum_{g=0}^\infty c_g   e^{- a_g T t } N^{2-2g}+ c_{\infty} e^{-N^2} ~,
\ee
special correlation functions are larger at late times.

One very special example is given by
$$ G_{\text{very special}}(t) \equiv \vev{ L_-(t) L_+(0)}_T .$$
Using $[H, L_-] = -{1\over R} L_- $ we have
$$ G_{\text{very special}}(t) = e^{- i t/R} \vev{ L_-(0) L_+(0)}_T $$
and hence
$$ \left| {G_{\text{very special}}(t) \over G_{\text{very special}}(0) } \right| = 1~; $$
the amplitude of this correlation does not decay at all.
This is analogous in hydrodynamics to correlations of the conserved quantities themselves, 
such as $ P^i(k=0)$, which are time independent.

There are other operators, in particular certain modes
of the stress-energy tensor, which can excite and destroy the oscillations
we have described, and therefore have non-decaying contributions
to their autocorrelation functions.
These receive oscillating contributions at one loop,
in close analogy with the calculation of \cite{CaronHuot:2009iq}, which finds
\be \vev{T^{xy}(t, k=0) T^{xy}(0, k=0)}_T  = T^2 \int_k \( G^{xtxt}(t,k) G^{ytyt}(t,k) 
+ G^{xtyt}(t,k) G^{xtyt}(t,k) \) \ee
where $G^{abcd}(t,k) \equiv \vev{T^{ab}(t, k) T^{cd}(0, -k)}_T$.
This is most easily understood (when a gravity dual is available) via the bulk one-loop Feynman diagram: 
\begin{fmffile}{fmfp23}  
 \unitlength=1mm  
\begin{fmfgraph}(32,20)   
\fmfleft{i1}
\fmfright{o1}
\fmf{curly}{i1,v1}
\fmflabel{$\ell=1$}{i1}
\fmflabel{gerbils}{v1}
\fmfdotn{v}{1}
\fmfdotn{v}{2}
\fmf{curly, left, tension=.1, label={$A_{t}(l=0)$}, l.side=left}{v1,v2}
\fmf{curly, right, tension=.1, label={$g_{tx}(l=0)$}, l.side=right}{v1,v2}
\fmf{phantom}{v1,v2}
\fmf{curly}{v2,o1}
\end{fmfgraph}
The contribution from $k=0$ in the momentum integral dominates at late times
and produces the power-law tail in $t$.

The analog in CFT on $S^{d-1}$ is given by the lowest angular-momentum mode of 
stress-energy tensor $\CO_\pi = \int_{S^{d-1}} T_{ij} \pi^{ij} $
where $\pi$ is a transverse traceless $J=2$ ($ \vec J = \vec L + \vec S$)
tensor spherical harmonic.  
The leading contribution to $\vev{\CO_\pi(t) \CO_\pi(0)}_T$ at late times 
can again be described by the one-loop diagram above.
The intermediate state now involves a sum over angular momenta rather than momenta;
the term where both of the intermediate gravitons sit in the oscillating mode
gives a contribution proportional to $e^{it/R}$ which does not decay;
all other contributions decay exponentially in time.

Similarly, if we have additional conserved global charges in our CFT, 
we can construct non-decaying contributions where the graviton
is in the oscillating mode and the bulk photon line carries the conserved charge, as follows:
\hskip.2in
\begin{fmfgraph}(32,20)   
\fmfleft{i1}
\fmfright{o1}
\fmf{wiggly}{i1,v1}
\fmflabel{$\ell=1$}{i1}
\fmflabel{gerbils}{v1}
\fmfdotn{v}{1}
\fmfdotn{v}{2}
\fmf{wiggly, left, tension=.2, label={$A_{t}(l=0)$}, l.side=top}{v1,v2}
\fmf{curly, right, tension=.1, label={$g_{tx}(l=0)$}, l.side=bottom}{v1,v2}
\fmf{phantom}{v1,v2}
\fmf{wiggly}{v2,o1}
\end{fmfgraph}
\end{fmffile}
This is analogous to the long-time tails in current-current correlators in hydrodynamics.

This paper 
\cite{Birmingham:2002ph}
observes oscillations in 
real-time correlation functions in finite-volume CFT in 1+1 dimensions.
Some of the explicit formulae are special to 1+1 dimensions.
Also, \cite{Maldacena:2001kr} presents some such correlators.

\section{Discussion}

We provide some context for our thinking about these oscillating states, which could have been studied
long ago\footnote{We are aware of the following related literature:
Recently, very similar states were used as ground-like states
for an implementation of an $AdS_2/CFT_1$ correspondence 
\cite{Chamon:2011xk}. Coherent states for $SL(2,\mathbb{R})$ were constructed in
\cite{Barut:1970qf}.
The states we study are not eigenstates of the lowering operator $L_-$,
but rather of linear combinations of  powers of $L_{+}$ and  $L_{-}$, and $H$. 
\cite{Chamon:2011xk} builds pseudocoherent states which are annihilated by a linear combination of $L_{-}$ and $H$.
Other early work which emphasizes the role of $SL(2,\mathbb{R})$ representation theory
in conformal quantum mechanics is
\cite{deAlfaro:1976je}.
}, and in particular for thinking about collective oscillations dual to bouncing black holes.

Our initial motivation was to consider whether it is always the case that the entanglement entropy of subregions in 
QFT grows monotonically in time.
If an entire system thermalizes, 
a subsystem should only thermalize faster,
since the rest of the system can behave as a thermal bath.

However, it is well-known that a massive particle in global AdS
oscillates about $\rho=0$ in coordinates where
\be
ds^2=R^2(-\cosh{\rho}^2dt^2+d\rho^2+\sinh{\rho}^2d\Omega^2),
\ee
with a period of oscillation $2\pi R$. This is because the geometry is a gravitational potential well.
Massive geodesics of different amplitudes of oscillation about $\rho=0$ are mapped to each other by isometries of AdS. Specifically, the static geodesic $\rho=0$ for all $t$ can be mapped to a geodesic oscillating about $\rho=0$ by a special conformal transformation.

The effect of a massive object in the bulk on the entanglement entropy of a subregion in the dual CFT 
is proportional to its mass in Planck units (see section 6 of \cite{Hubeny:2007xt}).
Only a very heavy object, whose mass is of order $N^2$, will affect the entanglement entropy
at leading order in the $1/N^2$ expansion
(which is the only bit of the entanglement entropy that we understand holographically so far).
 A localized object in $AdS$ whose mass is of order $N^2$ is a large black hole.
Therefore, 
by acting with an AdS isometry on a global AdS black hole,
one can obtain a state in the CFT in which the entanglement entropy oscillates
in time.



But this bouncing black hole is none other than the dual description of a collective oscillation built on thermal equilibrium, as introduced near  \eqref{eq:bb}. 
(Recall that 
in systems with a classical gravity dual, the thermal ensemble 
at temperature of order $N^0$ is dominated by energy eigenstates with energy of order $N^2$. )
The existence of this phenomenon is not a consequence of holography,
but rather merely of conformal invariance.

An important general goal is to clarify in which ways holographic CFTs are 
weird because of holography and in which ways they are weird just because they
are CFTs.  Our analysis demonstrates that 
these oscillations are an example of the latter.
The effect we have described arises
because of the organization of the CFT spectrum into towers 
of equally-spaced states.

In holographic calculations,
there is a strong temptation to 
study the global AdS extension
because it is geodesically complete.
This means compactifying the space on which the CFT lives
by adding the `point at infinity'.
We have shown here that this seemingly-innocuous 
addition can make a big difference for the late-time behavior!

\vskip.2in
We close with some comments and open questions.

\begin{enumerate}

\item
How are these oscillations deformed as we move away from the conformal fixed point? What does adding a relevant operator do to the oscillations? 
Such a relevant deformation should produce a finite damping rate 
for the mode.
This damping rate provides a new scaling function -- 
it has dimensions of energy and can depend only on $R$ 
and the coupling of the relevant perturbation $g$, 
and must vanish as $g\to 0$. 
If the scaling dimension of $g$ is $\nu$, the damping rate is 
$ \Gamma(g, R) = g^{1/\nu} \Phi(gR^\nu) $ 
where $ \Phi(x)$ is finite as $ x\to 0$.
It may be possible to determine this function $\Phi$ holographically.

\item
If we consider the special case of a CFT which is also a superconformal field theory,
there are other stable oscillations that we can make
by exponentiating the action of the fermionic symmetry generators $S_\pm^\alpha$.
Since $[H, S_\pm^\alpha] = \pm S_\pm^\alpha$, 
these modes have frequency $ {1\over 2R}$.
Fermionic exponentials are simple, and these states take the form
\be
\ket{\aaa(t)} = e^{\aaa_\alpha(t) S_+^\alpha } \ket{\Delta}
= \( 1 + \aaa_\alpha(t)  S_+^\alpha\) \ket{\Delta} 
\ee
with $\aaa(t) = e^{i t/2R} \aaa(0)$.

\item
We can define an analog 
for the bouncing black hole 
of the Aichelberg-Sexl shockwave 
\cite{Aichelburg:1970dh}
that results from a lightlike boost
a Schwarzchild black hole in flat space.
Here one takes the limit of lightlike boost
$\beta \to \infty$,
while simultaneously reducing the mass
to keep the energy fixed.
In this limit of the bouncing black hole, which merits further study, the profile of the {\it boundary}
energy density is localized on the wavefront.

\item 
For which background geometries $M$ can such states of CFT be constructed?
A sufficient condition is the existence of a CKV $\xi$ on $M$ whose Lie bracket
with the time-translation generator $\partial_t$ is of the form
$ [\partial_t,  \xi] = c \xi $ for some constant $c$.
It would be interesting to decide
whether there exist spacetimes $M$ with such CKVs 
where the constant $c$ remains finite as the volume of $M$ is taken to infinity
(unlike $S^{d-1}\times \mathbb{R}$ where $ c = R^{-1} \to 0$).
This would be interesting because in finite volume theories
do not thermalize anyway (at finite $N$) because
they are not in a thermodynamic limit.

\item 
It would be interesting to generalize these oscillations
to CFTs on spacetimes with boundary, with conformal boundary conditions.

\item 
The recent paper 
\cite{Dias:2011ss}
also identifies undamped oscillating states in holographic CFTs on $S^{2}$
using AdS gravity.
These `geons' are normalizible of classical AdS gravity
with frequency of oscillation $n/R$.
They clearly differ from the states constructed above in that 
they are excitations above the AdS vacuum, 
whereas our construction relies on broken conformal invariance.
There has been also some interesting recent 
work on {\it damped} oscillations in the approach to equilibrium 
in scalar collapse in AdS
\cite{Garfinkle:2011hm}.

\end{enumerate}

\vfill\eject
{\bf Acknowledgements}

We thank Ethan Dyer, Gary Horowitz, Roman Jackiw, Matthew Kleban, Lauren McGough, Mike Mulligan, Massimo Porrati, Steve Shenker, Lenny Susskind, Brian Swingle, Erik Tonni and Martin Zwierlein
for discussions, comments and encouragement.  This work was supported in part by
funds provided by the U.S. Department of Energy
(D.O.E.) under cooperative research agreement DE-FG0205ER41360,
and in part by the Alfred P. Sloan Foundation.

\begin{appendix}

\section{Conformal Algebra}
\label{app:algebra}

Take the coordinates of the flat spacetime on which the global conformal group $SO(2,d)$ acts linearly to be $(X_{-1},X_{0},X_{1},...,X_{d})$, with signature $(-, -, +,...,+)$. This is the embedding space of $AdS_{d+1}$ with boundary $\mathbb{R}\times S^{d-1}$. The compact subgroups $SO(2)$ and $SO(d)$ of $SO(2,d)$ correspond to time translation and spatial rotations in $\mathbb{R}\times S^{d-1}$, and are generated by rotations $J_{-10}$ and $\{J_{ij}\}$, $i,j=1,...,d$. In order to identify ladder operators acting on eigenstates of the Hamiltonian on $S^{d-1}$, $H=-J_{-10}$, we will situate $\{J_{\mu\nu}\}$ in an $so(1,d+1)$ algebra \cite{Minwalla:1997ka} with generators adapted to the global conformal symmetry of $\mathbb{R}^d$.\footnote{Our conventions are $J_{ab}=i(X_a\partial_b - X_b\partial_a)$, $[J_{ab},J_{cd}]=i(g_{ad}J_{bc}+g_{bc}J_{ad}-g_{ac}J_{bd}-g_{bd}J_{ac})$ for $a,b=-1,0,1,...,d$. The $so(1,d+1)$ algebra is manifest in the basis $J_{-10}^{'}=D^{'}, J_{ij}^{'}=J_{ij}, J_{-1i}^{'}=(P_{i}^{'}-K_{i}^{'})/2, J_{0i}^{'}=(P_{i}^{'}+K_{i}^{'})/2$ with signature $(-,+,+,...,+)$.}\footnote{Generators of $so(1,d+1)$ adapted to the conformal symmetry of $\mathbb{R}^{d}$, which we denote using primed letters $D', M'_{ij}, P'_{i}, K'_{i}$, can be related to generators of $so(2, d)$ adapted to the conformal symmetry of $\mathbb{R}^{1, d-1}$, $D, M_{ij}, P_{i}, K_{i}$, by expressing the latter generators in terms of rotations $J_{-10}$ and $J_{ij}$ in $AdS_{d+1}$. In particular, in the isomorphism between $\{D, M_{ij}, P_{i}, K_{i}\}$ and $\{J_{-10}, J_{ij}\}$, $J_{0-1}=\frac{1}{2}(P_0+K_0)$, from which we see $D'=i H = i J_{0 -1}=i \frac{1}{2}(P_{0}+K_{0})$.
}
\be
D'=-i J_{-10}= iH,~~
M_{ij}'=J_{ij},~~                            
P_{i}'=J_{i,-1}+iJ_{i0}\equiv L_{+}^{i},~~
K_{i}'=J_{i,-1}-iJ_{i0}\equiv L_{-}^{i}~~.
\ee


Note it was necessary to Euclideanize flat spacetime from $\mathbb{R}^{(1,d-1)}$ to $\mathbb{R}^{d}$ in order to identify the Hamiltonian in radially quantized $\mathbb{R}\times S^{d-1}$ with the dilation operator in flat spacetime. Now, from the familiar relations $[D^{'},P_{i}^{'}]=iP_{i}^{'}, [D^{'},K_{i}^{'}]=-iK_{i}^{'}$, and $[P_{i}^{'},K_{j}^{'}]=-2i(g_{ij}D^{'}-M_{ij}^{'})$ in $\mathbb{R}^{d}$, one can easily identify $d$ copies of the $SL(2,\mathbb{R})$ algebra with $H$ as the central operator,
\bea [H,L_{+}^{i}] &=&L_{+}^{i}, ~~~~[H,L_{-}^{i}]=-L_{-}^{i}, \cr 
[L_{+}^{i}, L_{-}^{j}]&=&2H \delta^{ij} + 2i M'^{ij}, ~~~~
[L_\pm^i, L_\pm^j]= 0 . \eea
Note the raising (lowering) operators commute with raising (lowering) operators, but raising operators do not commute with lowering operators.

\section{Norms of coherent states}
\label{app:norms}

Using formulas found in \cite{ban-optics}, 
we can determine 
the norm of a general coherent state of the form 
\eqref{eq:oscstate}.
For definiteness, will consider here 
coherent states built on energy eigenstates 
of the form $\ket{m, \epsilon_0}\propto L_+^m\ket{\epsilon_0}$ with $\ket{\epsilon_0}$ a primary state.
The norm-squared $|\CN|^{-2}$ of the state
$e^{\alpha L_+ + \beta L_-}\ket{m, \epsilon_0}$ is
\be
|\CN|^{-2} = \left(\left|\cosh{\sqrt{\alpha\beta}}\right|^2+\left|\frac{\beta}{\alpha}\right|\left|\sinh{\sqrt{\alpha\beta}}\right|^2\right)^{2(\epsilon+m)}\sum^{\infty}_{n=0}\frac{1}{n!^2}C^n\frac{(n+m)!}{m!}\frac{\Gamma(2\epsilon_0+n+m)}{\Gamma(2\epsilon_0+m)}
\ee
where
\begin{align*}
C=\cosh ^2\left(\sqrt{\alpha  \beta }\right) \left(\left|\frac{\beta  \tanh ^2\left(\sqrt{\alpha  \beta }\right)}{\alpha }\right|+1\right)^2\left(\frac{\sqrt{\frac{\beta }{\alpha }} \tanh \left(\sqrt{\alpha  \beta}\right)}{\left|\frac{\beta  \tanh ^2\left(\sqrt{\alpha \beta}\right)}{\alpha }\right|+1}+\frac{1}{2} \sqrt{\frac{\alpha }{\beta}}^{*} \sinh \left(2 \sqrt{\alpha  \beta}^{*}\right)\right)\\
\left(\frac{\sqrt{\frac{\beta }{\alpha }}^{*}\text{sech} ^2\left(\sqrt{\alpha  \beta }\right) \tanh \left(\sqrt{\alpha  \beta }^{*}\right)}{\left|\frac{\beta\tanh ^2\left(\sqrt{\alpha  \beta }\right)}{\alpha}\right|+1}+\sqrt{\frac{\alpha }{\beta }} \tanh \left(\sqrt{\alpha \beta }\right)\right).
\end{align*}

Using the ratio test for convergence, $\sum^{\infty}_{n=0} a_n$ converges if $\lim_{n\rightarrow\infty}\left|\frac{a_{n+1}}{a_n}\right|<1$, we get the condition for convergence that $|C|<1$. In the special case that  $\beta=-\alpha^*$, i.e. $\alpha L_+ + \beta L_-$ is anti-Hermitian or $e^{\alpha L_+ + \beta L_-}$ is unitary, $C=0$, so the norm squared is exactly $1$ as it should be. When $\beta = \alpha^*$, i.e. $\alpha L_+ + \beta L_-$  is Hermitian, the condition becomes $\sinh ^2\left(2|\alpha|\right)<1$.
When $ \beta = 0$, we find $C=|\alpha|^2$, and the condition for convergence is $ |\alpha|^2 <1$.

\section{Goldstone States}
\label{app:goldstone}

It is useful to interpret some of the states we have described by adapting Goldstone's theorem
\cite{Goldstone}.
The remarks in the following two paragraphs are useful in developing intuition for this adaptation,
but a reader impatient with discussion of holography
can skip to the holography-independent argument which follows.

In the Schwarzchild black hole in flat space, there is a static $\ell=1$ mode
\cite{Regge:1957td}
which has a very simple interpretation.
The black hole in flat space breaks translation invariance:
the $\ell=1$ mode is the Goldstone mode.
It just shifts the center of mass of the BH.

There is a strong analogy between 
the breaking of translation invariance by the Schwarzchild black hole in flat space
and
the breaking of conformal invariance by the Schwarzchild black hole in AdS.
But there is an important difference between momentum in flat space
and the conformal charges in AdS:  
unlike $[\vec p, H_{\text{flat}}] = 0$, the conformal charges do not commute with the Hamiltonian.
So there is a small modification of Goldstone's theorem which takes this into account and
leads to definite time dependence $e^{ - i t /R}$, rather than no time dependence.

More generally, the oscillations we construct in \eqref{eq:oscstate} and \eqref{eq:bb}
can be viewed as Goldstone states arising from the breaking of conformal symmetry
by the state on which the oscillation is built, the ``base state".
In fact there is such a mode for each of the $2d$ charges $Q^{\pm,i}$ in \eqref{eq:charges} 
which does not annihilate the base state.

The following algebraic argument shows that the state arising from spontaneous breaking of conformal symmetry associated with any of the charges $Q^{\pm,i}$ 
has frequency $\pm 1/R$, in agreement with the evolution found in \eqref{eq:evolution} for $n=1$.

To make the logic explicit, recall the usual Goldstone argument for a charge which commutes with $H$ 
in a relativistic QFT.
Proceed by noting that the broken current is an interpolating field for the Goldstone mode:
\be 
\label{eqn:goldstone1}
\bra{ \pi(k^\mu) } j^\mu(x) \ket{\text{symmetry-broken groundstate}} = i f_\pi k^\mu e^{-i kx}  ~.
\ee
Then current conservation gives
$
0 = \partial_\mu j^\mu \propto k_\mu k^\mu,
$
and hence the long-wavelength $\vec k = 0$ Goldstone mode $\ket{\pi}$ has $\omega=0$.

Here is the adaptation.
Let $Q$ be any of the charges $Q^{\pm,i}$. Significantly, $Q=Q(t)$ has explicit time-dependence. Let $\ket{\Delta}$ be a state of the CFT that breaks the conformal symmetry associated with $Q(t)$ but which is still 
stationary (the generalization to mixed states will be clear).
Let $\ket{\Delta, \pi (\omega)}$ be the Goldstone state expected from spontaneous symmetry breaking. Then we can parametrize the matrix element
\be \label{eq:gold}
\bra{\Delta, \pi(\omega)}Q(t)\ket{\Delta} = f_\pi e^{-i\omega t},
\ee
where $f_\pi$ is a constant that depends on the normalization of $Q(t)$,
and $\omega$ the frequency of the Goldstone excitation, which is to be determined.
Taking the partial derivative with respect to time on both sides,
\begin{align}
 - i \omega f_\pi e^{ - i \omega t }&= \bra{\Delta, \pi(\omega)} \partial_t Q(t) \ket{\Delta}\\
\nonumber&=  \bra{\Delta, \pi (\omega)} i [H, Q(t)] \ket{\Delta}\\
\nonumber&=\pm\frac{i}{R}\bra{\Delta, \pi (\omega)} Q(t) \ket{\Delta}=\pm \frac{i}{R}f_\pi e^{-i \omega t},
\end{align} 
where in the second line we have used ${d \over dt} Q = 0$. This shows $\omega =\pm 1/R$ for $Q^\pm$, as claimed.

\section{$\ell=1$ Mode in Global $AdS_{4}$-Schwarzchild Geometry}
\label{app:qnm}

Here we construct a linearized gravity mode of frequency $\omega = 1/R$ 
in the $AdS_4$-Schwarzchild black hole, whose existence and frequency are guaranteed by conformal symmetry, and which corresponds to a particular collective oscillation in the dual CFT. We study $AdS_4$ for definiteness, but the generalization to other dimensions 
should be clear. 
We proceed by finding a vector field $\xi$ 
in the spacetime which falls off too slowly at the $AdS$ boundary
to generate an equivalence of configurations,
but falls off quickly enough to produce a normalizable metric perturbation
\be
\label{eq:gauge}
h_{ab} = \xi_{a;b} + \xi_{b;a}~~.
\ee
By the correspondence with flat-space Schwarzchild described in \ref{app:goldstone},
this mode is analogous to the Goldstone
mode for broken translation invariance, 
\ie~the mode that translates the center-of-mass of the black hole.

We demand that near the $AdS$ boundary (at $r=\infty$ in the coordinates we'll use here) 
$\xi$ approaches a conformal isometry;
in the empty global $AdS_4$ background 
\be
ds^2=-f_0(r)dt^2+\frac{1}{f_0(r)}dr^2+r^2(d\theta^2+\sin{\theta}^2d\phi^2)~,~~ f_0(r)=1+\frac{r^2}{R^2}~,
\ee
one $\ell=1, m_z = 0$ KV takes the form \cite{Henneaux:1985tv}
\be
\label{eq:ckv}
\xi_0 =  
e^{-it/R}\(\frac{i r \cos (\theta)}{\sqrt{1+\frac{r^2}{R^2}}}{\partial\over\partial t}-R \cos{\theta}\sqrt{1+\frac{r^2}{R^2}}{\partial\over\partial r}+ \frac{R}{r}\sin{\theta}\sqrt{1+\frac{r^2}{R^2}}{\partial\over\partial \theta}\)~.
\footnote{The other $m_z=0$ KV is obtained by complex conjugation, and $m_z = \pm 1 $ KVs can be constructed using the Lie bracket
with $so(3)$ raising and lowering operators 
\be
\nonumber
M^\pm = 
e^{ \pm i \varphi } \( {\partial\over \partial\theta} \pm 
 i \cot \theta {\partial \over \partial \varphi} \)~.
\ee}
\ee
%
The $AdS_4$-Schwarschild metric is
\be
ds^2 = - f(r) dt^2 + {dr^2\over f(r)} + r^2 \( d\theta^2 + \sin^2\theta d\varphi^2 \), ~~
f(r) = 1+ \frac{r^2}{R^2}  - { r_+ \over r }~.
\ee

We make the ansatz 
\be
\xi=e^{-i \omega t}\(i g(r) \cos{\theta}{\partial\over\partial t}+R~h(r) \cos{\theta}{\partial \over \partial r}+R~j(r)\sin{\theta}{\partial \over \partial \theta}\)~~.
\ee
We demand that the resulting metric perturbation $h_{ab}$ 
\begin{itemize}
\item is in the `gaussian normal' gauge $h_{ra}=0$ customary for holography, 
\item is normalizable, corresponding to a state in the dual CFT, 
\item satisfies the b.c. $\xi^a \rightarrow \xi_0^a$ as $r\rightarrow \infty$.
\end{itemize}
Our boundary condition $\xi \buildrel{r\to\infty} \over {\to} \xi_0$ determines $\omega=1$,
but leaving $\omega$ arbitrary provides a check.

Imposing the gaussian normal gauge $h_{ra}=0$, we find
\begin{align}
\label{eq:presol}
h(r)&=c_1 \sqrt{f(r)}~,\\
\nonumber g(r)&=c_1\int^{\infty}_{r} {dr'\over f(r')^{3/2}} + c_2~,\\
\nonumber j(r)&=-c_1\int^{\infty}_{r}{dr'\over r'^2 f(r')^{1/2}}+c_3~.
\end{align}
Normalizability determines $\omega^2=1$.
Demanding $\xi \rightarrow \xi_0$ as $r\rightarrow \infty$ gives
\be
c_1=-1~,~ c_2=R~,~ c_3=1/R~,
\ee
after which one can check that the normalizability conditions $h_{tt},~h_{t\theta},~h_{\theta\theta},~h_{\phi\phi}
\buildrel{r\to\infty}\over{\sim} O({1\over r})$ 
are satisfied. 

The resulting metric perturbation $h_{ab}$ is 
\bea
h_{ab} dx^a dx^b &=& 
e^{-\frac{i t}{R}}
\[
-\cos\theta \left(2 f g+h \left(2 r+{R^2r_+\over r^2} \) \) 
dt^2 
+  i \sin\theta \left(f g-r^2
   j\right) dt d\theta
   \right.
\cr\cr
&&\left.
+ 2 r R   (h+r j) \cos\theta\( d \theta^2
+  \sin^2\theta d\varphi^2\) 
\]~.
\eea

\section{Oscillating Observables for a Bouncing Black Hole}
\label{app:bbh}

For simplicity, we work with the non-rotating $2+1$-dimensional BTZ black hole, using coordinates in which its metric is
\be
\label{eq:metric}
ds^2=-(\frac{r^2}{R^2}-\frac{r_{+}^2}{R^2})dt^2+\frac{dr^2}{\frac{r^2}{R^2}-\frac{r_{+}^2}{R^2}}+\frac{r^2}{R^2}dx^2.
\ee
Here $R$ is the AdS radius, $r_{+}$ the radius of the black hole, and $\frac{x}{R}\sim \frac{x}{R} + 2 \pi$.

The stress-energy tensor in the finite-temperature CFT corresponding to this large global AdS black hole can be obtained by varying the bulk Einstein action with respect to the boundary metric and using local counterterms \cite{Balasubramanian:1999re}. With light-like coordinates $x_{\pm}=t\pm x$,
\be
T_{++}=T_{--}=\frac{r_{+}^2}{32 \pi R^3}
\ee
and $T^{\mu}_{\mu}\propto \mathcal{R}=0$, where $\mathcal{R}$ is the scalar curvature of the boundary metric. 

The stress-energy tensor for a BTZ black hole, after a coordinate transformation corresponding to a boost in $SO(2,2)$, is then obtained most easily by transforming $T_{++}$ and $T_{--}$ under the conformal transformation induced on the boundary by the $SO(2,2)$ boost. Note $T^{\mu}_{\mu}$ remains zero. For the boost $e^{i \beta J^{01}}$, $v=\tanh{\beta}$, acting on embedding coordinates of $AdS_3$, the corresponding coordinate transformation on coordinates $(t, x, r)$ in \eqref{eq:metric} is given by
\begin{align}
\label{eq:trans}
\tan{\frac{t'}{R}}&=\frac{\sqrt{1-v^2}\sqrt{1+(\frac{r}{R})^2}\sin{\frac{t}{R}}}{\sqrt{1+(\frac{r}{R})^2}\cos{\frac{t}{R}}+v\frac{r}{R}\cos{\frac{x}{R}}},\\
\tan{\frac{x'}{R}}&=\frac{\sqrt{1-v^2}\frac{r}{R}\sin{\frac{x}{R}}}{v\sqrt{1+(\frac{r}{R})^2}\cos{\frac{t}{R}}+\frac{r}{R}\cos{\frac{x}{R}}},\nonumber\\
\frac{r'}{R}&=\sqrt{\left(\frac{v}{\sqrt{1-v^2}}\sqrt{1+\left(\frac{r}{R}\right)^2}\cos{\frac{t}{R}}+\frac{1}{\sqrt{1-v^2}}\frac{r}{R}\cos{\frac{x}{R}}\right)^2+\left(\frac{r}{R}\sin{\frac{x}{R}}\right)^2},\nonumber
\end{align}

\begin{figure}[h] \begin{center}
\includegraphics[height=50mm, width=65mm]{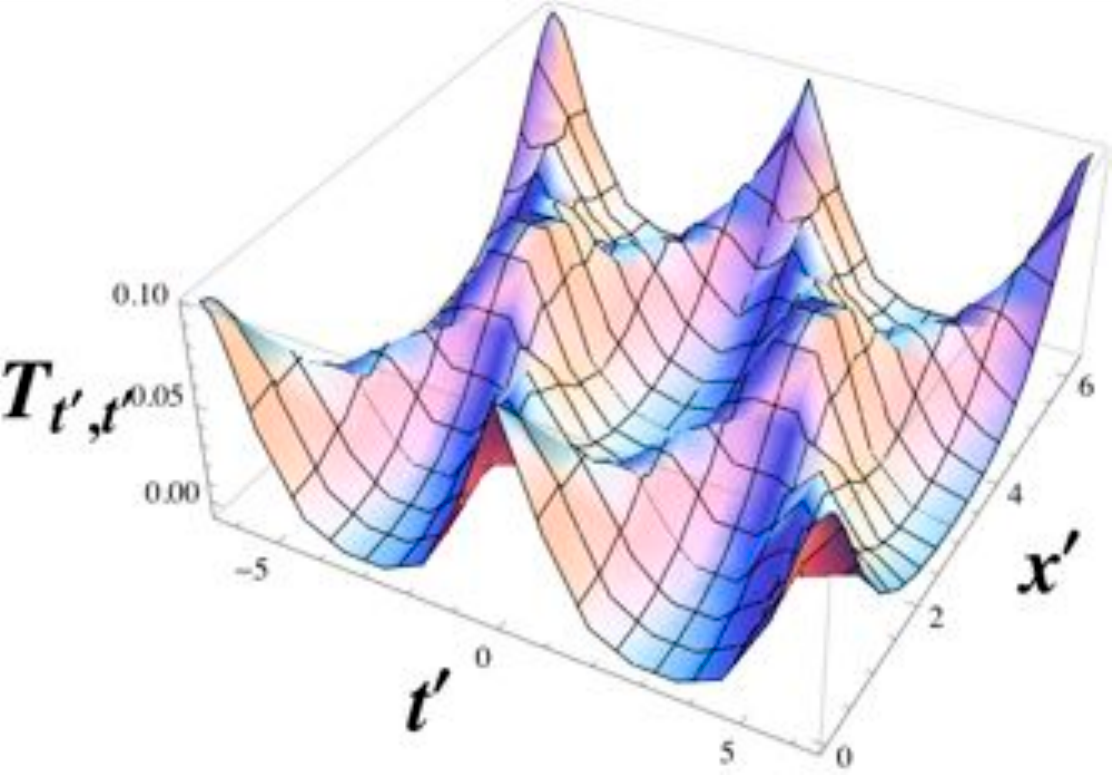}
\hskip.4in
\includegraphics[height=50mm, width=65mm]{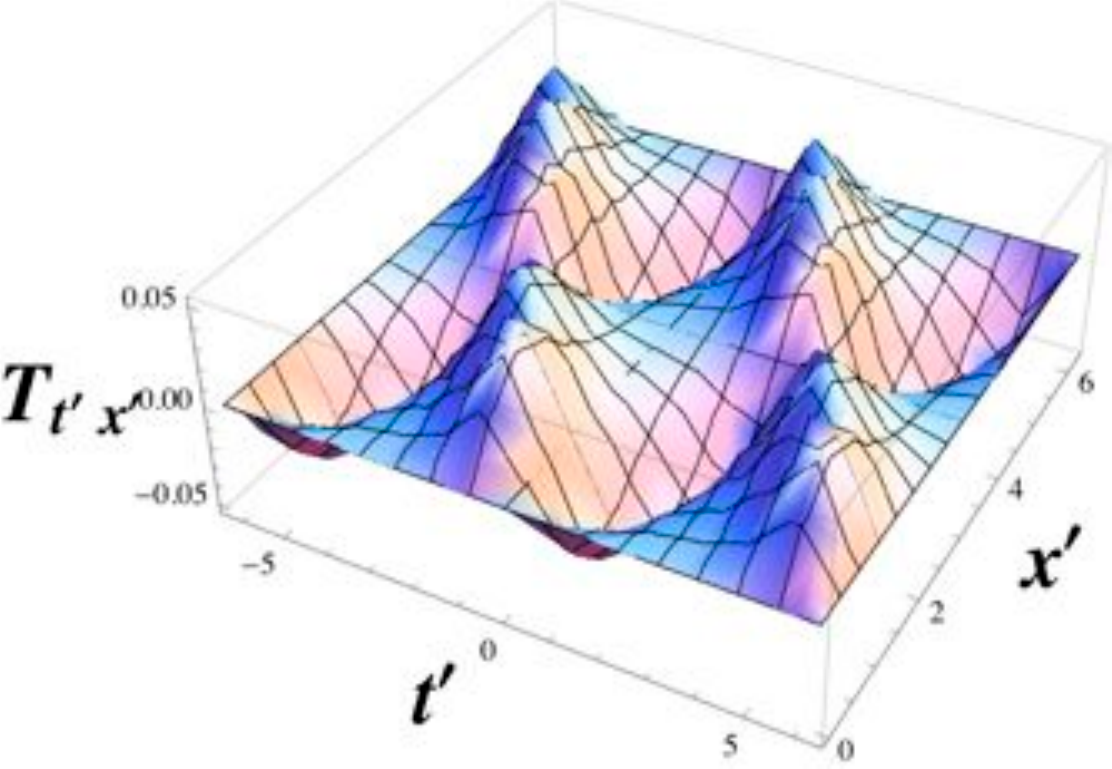}
\end{center}
\caption{\label{fig:stress}
Energy and momentum density as a function of position $x'$ and time $t'$ with $R=r_{+}=G=1$, $v=0.5$.
}
\end{figure}

\begin{figure}[h] \begin{center}
\includegraphics[height=40mm, width=65mm]{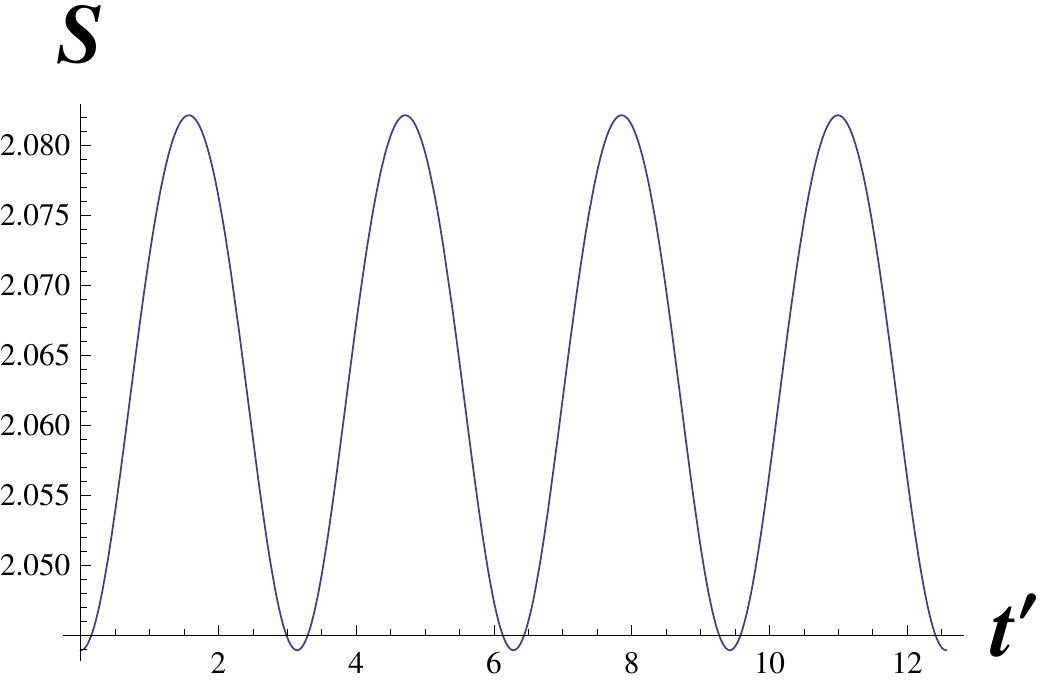}
\hskip.4in
\includegraphics[height=40mm, width=65mm]{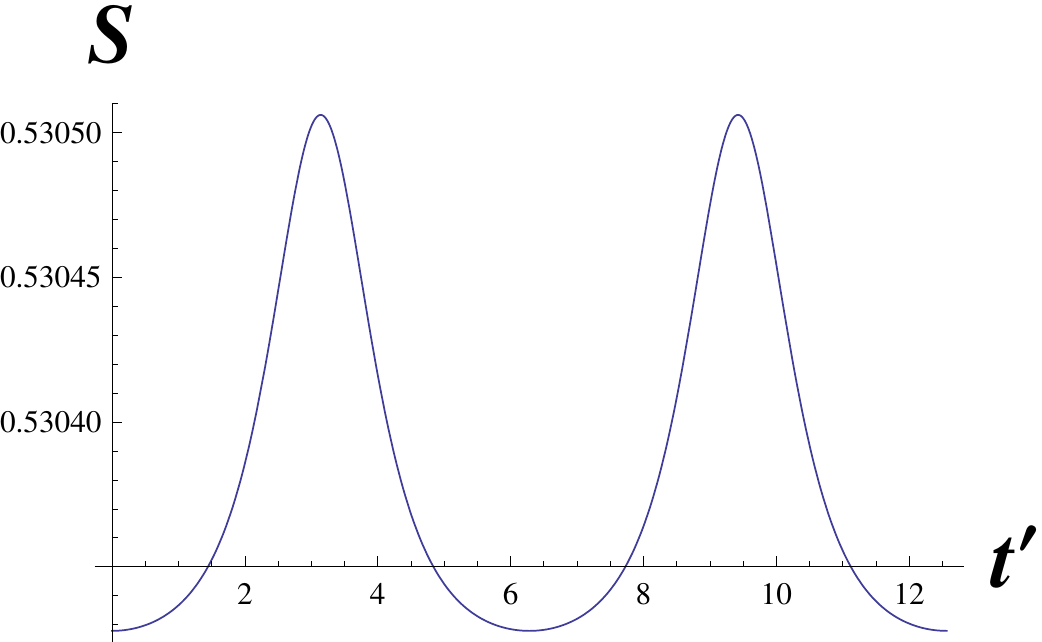}
\end{center}
\caption{\label{fig:entropy}
Entanglement entropy of the interval $0\leq x' \leq \pi$, left, and $0\leq x' \leq \pi/64$ with $R=r_{+}=G=c=1$, $v=0.5$, and cutoff $\epsilon=0.01$.}
\end{figure}

from which follows
\begin{align}
T_{t't'}=T_{x'x'}&=\frac{\frac{\left(1-v^2\right) \left(R^2+r_{+}^{2}\right)}{\left(v \cos \left(\frac{t'-x'}{R}\right)-1\right)^2}+\frac{\left(1-v^2\right) \left(R^2+r_{+}^2\right)}{\left(v \cos \left(\frac{t'+x'}{R}\right)-1\right)^2}-2 R^2}{32 \pi  G R^3},\\
T_{t'x'}&=\frac{v \left(1-v^2\right) \left(R^2+r_{+}^2\right) \sin \left(\frac{t'}{R}\right) \sin \left(\frac{x'}{R}\right) \left(v \cos \left(\frac{t'}{R}\right) \cos \left(\frac{x'}{R}\right)-1\right)}{8 \pi  G R^3\left(v \cos \left(\frac{t'-x'}{R}\right)-1\right)^2 \left(v \cos \left(\frac{t'+x'}{R}\right)-1\right)^2}.
\end{align}

Note $J^{01}=(L_{+}^{1}-L_{-}^{1})/(2i)$, so that the stress-energy tensor above is that of the collective oscillation in \eqref{eq:bb} with $\alpha=\beta$. It indeed manifestly oscillates in time. Its two nonzero components are plotted in Fig.~\ref{fig:stress}.

We can also confirm that in the same state, the entanglement entropy calculated by the covariant holographic method in \cite{Hubeny:2007xt}, of a spatial subregion with respect to the rest of $S^{1}$, oscillates.

Given the endpoints $x_1', x_2'$ of such a spatial subregion, its entanglement entropy as a function of $t'$ is given by 
\be
\label{eq:ee}
S(x_1', x_2',t')=\frac{c}{6} \log{L(x_1',x_2',t')},
\ee
where $L(x_1',x_2',t')$ is the length of the space-like geodesic ending at points $p_1'=(t', x_1', r_c'), p_2'=(t', x_2', r_c')$, with $r_c'=\frac{1}{\epsilon}$ an infrared cutoff in the $SO(2,2)$-boosted BTZ black hole. 

The desired space-like geodesic may be obtained by first mapping $p_{i}'$ to $p_{i}=(t_{i}, x_{i}, r_{i})$, $i=1,2$, where $(t,x,r)$ are non-boosted coordinates given by the inverse coordinates transformation of \eqref{eq:trans}, and again mapping $p_{i}$ to $q_{i}=(w_{+\, i}, w_{-\, i}, z_{i})$, $i=1,2$, where $(w_{+},w_{-}, z)$ are coordinates in which the BTZ black hole has the manifestly AdS metric
\be
ds^2=R^2\left(\frac{dw_{+}dw_{-}+dz^2}{z^2}\right).
\ee
The coordinate transformation from $(t,x,r)$ to $(w_{+},w_{-},r)$ is given by
\be
w_{\pm}\equiv X \pm T=\frac{\sqrt{r^2-r_{+}^2}}{r}e^{\frac{(x\pm t)r_{+}}{R^2}}, z=\frac{r_{+}}{r}e^{\frac{x r_{+}}{R^2}}.
\ee

In the newest coordinates, the space-like geodesic with endpoints $q_{i}=(w_{+\, i}, w_{-\, i}, z_{i})$, $i=1,2$ can be boosted by the mapping $w_{\pm}'=\gamma^{\pm 1} w_{\pm}$, $\gamma=\sqrt{\frac{1-\beta}{1+\beta}}$, with $\beta$ the usual Lorentz boost parameter in coordinates $(T,X)$, to lie on a constant $T$ hypersurface. The resulting geodesic is a circular arc satisfying 
\be
(\frac{\gamma w_{+}+\gamma^{-1}w_{-}}{2}-A)^2+z^2=B^2,\\
\frac{\gamma w_{+}-\gamma^{-1}w_{-}}{2}=C,
\ee
where constants $\gamma, A, B, C$ can be determined by the two endpoints $q_{i}$, $i=1,2$, and which has length
\be
L=R \log{\frac{(w_{+\, 2}-w_{+\, 1})^2(w_{-\, 2}-w_{-\, 1})^2+2(w_{+\, 2}-w_{+\, 1})(w_{-\, 2}-w_{-\, 1})(z_1^2+z_2^2)+(z_2-z_1)^2}{(w_{+\, 2}-w_{+\, 1})(w_{-\, 2}-w_{-\, 1})z_1z_2}}.
\ee

Translating back to original coordinates $(t', x', r')$, $L=L(x_1', x_2',t')$, one has the holographic entanglement entropy \eqref{eq:ee} in a bouncing black hole geometry. The smoothly oscillating entanglement entropy is plotted for two different intervals in the spatial domain $S^1$ of the CFT in Fig.~\ref{fig:entropy}.

\section{Oscillations in Galilean Conformal Field Theory}
\label{app:galilean}

We point out that these oscillations can be observed in experiments on ultracold fermionic atoms at unitarity, 
and that related modes have already been studied in detail
\cite{Kinast, Turlapov, grimm, Cao2011} (for reviews
of the subject see
\cite{zwierlien,zwerger,stringarireview,castinreview}).  
The specific mode we discuss
has been predicted previously by very different means in \cite{Castin2004}.

In the experiments, lithium atoms are cooled in an optical trap 
and their short-ranged two-body interactions are tuned 
to a Feshbach resonance via an external magnetic field.
Above the superfluid transition temperature, 
this physical system is described by a Galilean-invariant CFT \cite{Werner, Nishida:2007pj}.
The symmetries of such a system 
comprise a Schr\"odinger algebra, which importantly for our purposes 
contains a special conformal generator $C$.
(This symmetry algebra has been realized holographically
by isometries in 
\cite{Son:2008ye, Balasubramanian:2008dm}
(see also \cite{Duval:1990hj})
and more generally in \cite{Balasubramanian:2010uw}.)
The Hamiltonian for such a system in a spherically-symmetric harmonic trap 
$H_{\text{osc}} $
is related to the free-space Hamiltonian $H$ 
by \cite{Werner, Nishida:2007pj}
\be H_{\text{osc}} = H + \omega_0^2 C .\ee
This $H_{\text{osc}}$ is analogous to the 
Hamiltonian of relativistic CFT on the sphere
in that its spectrum is determined by the spectrum of scaling dimensions of operators.

In the experiments of \cite{Kinast, Turlapov}, ``breathing modes" of the fluid were excited by varying the frequency of the trap.  One goal was to measure the shear viscosity of the strongly-coupled fluid 
(for a useful discussion, see \S5.2 of \cite{Schafer:2009dj}).  
Energy is dissipated via shear viscosity in these experiments because the trapping potential 
is not isotropic. 
The anisotropy of the trap breaks the special conformal generator.
If the trap were spherical, 
our analysis would apply, and we predict that the mode 
with frequency $2 \omega_0$ would not be damped,
to the extent that our description is applicable
(\eg~the trap is harmonic and spherical and the coupling to the environment can be ignored)\footnote
{Note that this is not the lowest-frequency mode of the spherical trap;
linearized hydrodynamic analysis predicts a linear mode with frequency $ \sqrt{2} \omega_0$.}.

This prediction is consistent with the linearized hydro analysis of \cite{heiselberg, stringari2004},
and one can check that the sources of dissipation included in
\cite{Cao2011, Schafer:2009dj} all vanish for the lowest spherical breathing mode.
The mode we predict is adiabatically connected to the breathing mode studied in 
\cite{Kinast, Turlapov, Cao2011, grimm}.
Note that an infinite-lifetime mode of frequency $2 \omega_0$ is also 
a prediction for a free non-relativistic gas.
Indeed, this is also a Galilean CFT, though a much more trivial one.

There is a large (theoretical and experimental) literature on the collective modes of trapped quantum gases, 
\eg~\cite{heiselberg, stringari2004, stringari2005, stringari2007}.
Much of the analysis of these collective modes in the literature relies on
a hydrodynamic approximation.
In this specific context of unitary fermions in 2+1 dimensions, 
{\it linearized} modes of this nature were described 
in a full quantum mechanical treatment at zero temperature
by Pitaevskii and Rosch \cite{Pitaevskii}
and further studied in 3+1 dimensions by Werner and Castin \cite{Werner}.  
Further, their existence was attributed to a hidden so$(2,1)$ symmetry of the problem,
which is the relevant part of the Schr\"odinger symmetry.
The undamped nonlinear mode at $2 \omega_0$ was described in \cite{Castin2004}.
Here we make several additional points: 
\begin{itemize}
\item There are such stable modes at any even multiple of the frequency of the harmonic potential.
\item The fully-nonlinear modes of finite amplitude can be explicitly constructed, and remain undamped.
\item These modes are superuniversal -- they can be generalized to oscillations in any conformal field theory.
\end{itemize}

In a Galilean CFT with a harmonic potential, the oscillations can be constructed as follows.
Consider an eigenstate of $H_{osc}=H+ \omega_0^2 C$ constructed from a primary operator $\CO$ with dimension $\Delta_{\CO}$ \cite{Nishida:2007pj},
\be\ket{\Delta_\CO} = e^{-H} \CO^\dagger \ket{0},\;H_{osc}\ket{\Delta_\CO}=\Delta_{\CO}\ket{\Delta_{\CO}}. \ee
Defining ladder operators $L_\pm \equiv H-\omega_0^2 C \pm  i \omega_0 D$, the states 
\be\label{eq:galosc} \exp{\( {\alpha_{0} L_{+}+\beta_{0} L_{-}} \)}\ket{\Delta_{\CO}}\ee
with $\alpha_{0}$, $\beta_{0}$ c-numbers, evolve under $H_{osc}$ as 
\be  e^{- i t \Delta_\CO}\exp{\({e^{-2 i \omega_0 t }\alpha_0 L_{+}+e^{2 i \omega_0 t}\beta_{0}L_{-} }\)} \ket{\Delta_\CO}.\ee
The algebraic manipulations which demonstrate this time evolution are identical to those for coherent states of a simple harmonic oscillator. 

The Galilean generalization of \eqref {eq:general}
should be clear.
The crucial property of the CFT spectrum is again the existence of equally-spaced levels connected to $\ket{\Delta_{\CO}}$ by the ladder operators $L_+, L_{-}$.

Any real trap will be slightly anisotropic.
Following \cite{Cao2011, Schafer:2009dj}, estimates can be made using linearized hydrodynamics for the damping rate 
arising from the resulting shear of the fluid, in terms of the measured shear viscosity (see 
eqn.~(159) of \cite{Schafer:2009dj}).  We are not prepared to estimate other sources of dissipation.
It would be interesting to use softly-broken conformal invariance to predict the frequencies 
and damping rates of collective modes in slightly anisotropic traps in the nonlinear regime.

\end{appendix}

\bibliography{undamped}
\end{document}

\end{document}